\documentclass[11pt]{article}
\usepackage[utf8]{inputenc}
\usepackage[T1]{fontenc}
\usepackage{multicol}
\usepackage{authblk}
\usepackage{amsmath,amssymb,amsfonts,amsthm}
\usepackage[top=.80in, bottom=.80in, left=.80in, right=.80in]{geometry}
\newcommand{\Keywords}[1]{\par\noindent
{\small{\em Keywords\/}: #1}}

\catcode`,\active

\catcode`\,12

\newcommand*\pFqskip{8mu}
\catcode`,\active
\newcommand*\pFq{\begingroup
        \catcode`\,\active
        \def ,{\mskip\pFqskip\relax}%
        \dopFq
}
\catcode`\,12
\def\dopFq#1#2#3#4#5{%
        {}_{#1}F_{#2}\biggl[\genfrac..{0pt}{}{#3}{#4};#5\biggr]%
        \endgroup
}

\newcommand{\class}[1]{\par\noindent
{\small{AMS classification scheme numbers\/}: #1}}
\newcommand{\pd}{\partial}

\newcommand{\under}[1]{_{#1}}
\numberwithin{equation}{section}
\title{The equitable Racah algebra from three $\mathfrak{su}(1,1)$ algebras}
\date{}
\author[1]{Vincent X. Genest\thanks{genestvi@crm.umontreal.ca}}
\author[1]{Luc Vinet}
\author[2]{Alexei Zhedanov}
\affil[1]{Centre de Recherches Math\'ematiques, Universit\'e de Montr\'eal, C.P. 6128, Succursale Centre-ville, Montr\'eal, Qu\'ebec, H3C 3J7, Canada}
\affil[2]{Donetsk Institute for Physics and Technology, Donetsk 83114, Ukraine}
\begin{document}
\maketitle
\thispagestyle{empty}
\hrule
\begin{abstract}
\noindent
The Racah algebra, a quadratic algebra with two independent generators, is central in the analysis of superintegrable models and encodes the properties of the Racah polynomials. It is the algebraic structure behind the $\mathfrak{su}(1,1)$ Racah problem as it is realized by the intermediate Casimir operators arising in the addition of three irreducible $\mathfrak{su}(1,1)$ representations. It has been shown that this Racah algebra can also be obtained from quadratic elements in the enveloping algebra of $\mathfrak{su}(2)$. The correspondence between these two realizations is here explained and made explicit.
\\

\\
\class{17A45,\,17B35,\,17B80}
\end{abstract}
\hrule
\section{Introduction}
The Racah algebra, which connects superintegrable models to Racah polynomials \cite{Granovskii-1992-07,Kalnins-2007-09}, is more and more understood to have a universal role \cite{Genest-2013-tmp-1}. The main objective of this paper is to show that the equitable presentation of the Racah algebra emerges when the addition of three $\mathfrak{su}(1,1)$ representations of the positive-discrete series is considered and furthermore, to establish the relation that this framework has with the one in which the equitable presentation is obtained from quadratic elements in the universal enveloping algebra of $\mathfrak{su}(2)$ \cite{Gao-2013-06}. This will be done using Bargmann realizations by reducing the 3-variable model of the 3-summand $\mathfrak{su}(1,1)$ representation to the 1-variable realization of the Racah algebra stemming from the standard representation of $\mathfrak{su}(2)$ on holomorphic functions.

\subsection{Racah algebra}
The Racah algebra is the most general quadratic algebra with two algebraically independent generators, say $A$ and $B$, which possesses representations with ladder relations \cite{Granovskii-1989}. Upon introducing an additional generator $C$ defined by
\begin{align*}
[A,B]=C,
\end{align*}
where $[x,y]=xy-yx$, the Racah algebra is characterized by the commutation relations
\begin{subequations}
\label{Def-Rel}
\begin{align}
[B,C]&=B^2+\{A,B\}+d\,B+e_1,
\\
[C,A]&=A^2+\{A,B\}+d\,A+e_2,
\end{align}
\end{subequations}
where $\{x,y\}=xy+yx$. The Casimir operator, which commutes with all the generators of the Racah algebra, has the expression \cite{Granovskii-1988-10}
\begin{align}
\label{Casimir-Racah}
Q=\{A^2,B\}+\{A,B^2\}+A^2+B^2+C^2+(d+1)\{A,B\}+(2e_1+d)A+(2e_2+d)B.
\end{align}
In the realization that we shall be using, the parameters $d$, $e_1$, $e_2$ are expressed in terms of four real parameters $\lambda_i$, $i=1,\ldots,4$, as follows:
\begin{align*}
d=(\lambda_1+\lambda_2+\lambda_3+\lambda_4)/2,\quad e_1=(\lambda_1-\lambda_4)(\lambda_2-\lambda_3)/4,\quad e_2=(\lambda_1-\lambda_2)(\lambda_4-\lambda_3)/4.
\end{align*}
Note that the Racah algebra \eqref{Def-Rel} is invariant under the duality transformation $A\leftrightarrow B$, $e_1\leftrightarrow e_2$. 

The Racah algebra is intimately related to the Racah polynomials \cite{Genest-2013-tmp-1,Granovskii-1988-10,Granovskii-1989}, which sit atop the discrete part of the Askey scheme of hypergeometric orthogonal polynomials \cite{Koekoek-2010}. This relation emerges, on the one hand, from the representation theory of the Racah algebra. Indeed, finite-dimensional irreducible representations of \eqref{Def-Rel} can be obtained in bases where either $A$ or $B$ is represented by a diagonal matrix. In these representations, the non-diagonal generator is tridiagonal, which means that $A$ and $B$ realize a Leonard pair \cite{Terwilliger-2001-06}. In this picture, the Racah polynomials arise as the expansion coefficients between the eigenbases respectively associated to the diagonalization of $A$ and $B$. On the other hand, one can conversely arrive at the Racah algebra from the bispectrality properties of the Racah polynomials \cite{Genest-2013-tmp-1}. Indeed, upon identifying $A$ with the recurrence operator (viewed as multiplication by the variable) and taking $B$ as the difference operator of the Racah polynomials, it is checked that the defining relations \eqref{Def-Rel} are satisfied with values of the algebra parameters related to those of the Racah polynomials. 

Quite significantly, the Racah algebra has been found to be the symmetry algebra of the generic superintegrable 3-parameter system on the 2-sphere \cite{Genest-2013-tmp-1,Kalnins-2007-09}. This explains why the overlap coefficients between wavefunctions separated in different spherical coordinate systems are given in terms of Racah polynomials. Moreover, since it has been shown in \cite{Kalnins-2013-05} that all superintegrable systems in two dimensions with constants of motion of degree not higher than two in momenta are limits or specializations of the generic model on the 2-sphere, it follows that the symmetry algebras of these problems can all be obtained as special cases or contractions of the Racah algebra. Note that the Racah polynomials also arise in the interbasis expansion coefficients for the isotropic oscillator in the three-dimensional space of constant positive curvature \cite{Hakobyan-1999-07}.

Another manifestation of the Racah algebra is in the context of the Racah problem for both the $\mathfrak{su}(2)$ and the $\mathfrak{su}(1,1)$ Lie algebras \cite{Granovskii-1988-10}. The $\mathfrak{su}(1,1)$ case will be reviewed below. In considering the addition of 3 representations of $\mathfrak{su}(1,1)$ from the positive-discrete series, it shall seen that the intermediate Casimir operators associated to pairs of representations do satisfy the defining relations \eqref{Def-Rel}. In \cite{Genest-2013-tmp-1}, this observation has been related to the determination of the symmetry algebra of the aforementioned superintegrable model on the 2-sphere.

\subsection{Equitable presentation of the Racah algebra}
It is possible to exhibit a $\mathbb{Z}_3$-symmetric or equitable presentation of the Racah algebra \cite{Gao-2013-06,Kalnins-2000-04}. To that end, one first defines $X$, $Y$, $\Omega$ from $A$, $B$, $C$ as follows:
\begin{align}
\label{XYZ-2}
X=-2A-\lambda_1,\qquad Y=-2B-\lambda_2,\qquad \Omega=2C,
\end{align}
and also introduces a generator $Z$ related to $X$ and $Y$ by the relation
\begin{align}
\label{Sum}
X+Y+Z=\lambda_4.
\end{align}
It follows from \eqref{XYZ-2}, \eqref{Sum} and the definition of $C$ that
\begin{align*}
[X,Y]=[Y,Z]=[Z,X]=2\Omega.
\end{align*}
Rewriting the relations \eqref{Def-Rel} in terms of $X$, $Y$ and $Z$, one easily obtains
\begin{subequations}
\label{Z3-Symmetric-Racah}
\begin{align}
[X,\Omega]&=YX-XZ+(\lambda_1-\lambda_2+\lambda_3)\,Y-(\lambda_1+\lambda_2-\lambda_3)\,Z+f_1,
\\
[Y,\Omega]&=ZY-YX+(\lambda_2-\lambda_3+\lambda_1)\,Z-(\lambda_2+\lambda_3-\lambda_1)\,X+f_2,
\\
[Z,\Omega]&=XZ-ZY+(\lambda_3-\lambda_1+\lambda_2)\,X-(\lambda_3+\lambda_1-\lambda_2)\,Y+f_3,
\end{align}
\end{subequations}
where the structure parameters $f_1$, $f_2$, $f_3$ have the expression
\begin{align*}
f_1&=\big[\lambda_1(\lambda_2+\lambda_4)+\lambda_3(\lambda_2-\lambda_4)-2\lambda_1\lambda_3\big],
\\
f_2&=\big[\lambda_2(\lambda_3+\lambda_4)+\lambda_1(\lambda_3-\lambda_4)-2\lambda_2\lambda_1\big],
\\
f_3&=\big[\lambda_3(\lambda_1+\lambda_4)+\lambda_2(\lambda_1-\lambda_4)-2\lambda_3\lambda_2\big].
\end{align*}
The commutations relations \eqref{Z3-Symmetric-Racah} are manifestly invariant under cyclic permutations of $(X,Y,Z)$ and $(\lambda_1,\lambda_2,\lambda_3)$ and are referred to as the $\mathbb{Z}_3$-symmetric Racah relations. Given  \eqref{Sum}, they are obviously equivalent to \eqref{Def-Rel}. Recently, it has been shown that these equitable relations are realized by quadratic elements in $\mathcal{U}(\mathfrak{su}(2))$ \cite{Gao-2013-06}. It is the purpose of this paper to establish the correspondence between this last realization of $X$, $Y$, $Z$ and the Casimir operators of the $\mathfrak{su}(1,1)$ Racah problem.

\subsection{Outline}
The outline of the paper is as follows. In Section 2, we review the Racah problem for $\mathfrak{su}(1,1)$. We introduce the intermediate Casimir operators, record the relation between them and recall the definition of $6j$-symbols as coefficients between eigenbases corresponding to the diagonalization of two intermediate Casimir operators. In Section 3, we show that the intermediate Casimir operators associated to the $\mathfrak{su}(1,1)$ Racah problem realize the Racah algebra. In Section 4, we consider the Racah problem in the Bargmann picture and show how it reduces to the determination of the overlap coefficients between solutions to pairs of hypergeometric Strum-Liouville problems. The connection with the realization of the Racah algebra in terms of 3 (linearly related) quadratic elements in $\mathcal{U}(\mathfrak{su}(2))$ is then completed in Section 5 where it is shown that the reduction of the intermediate Casimir operators to hypergeometric Sturm--Liouville operators allows an identification with the quadratic elements in the Bargmann realizations of $\mathfrak{su}(2)$.
\section{The $\mathfrak{su}(1,1)$ Racah problem}
In this section, the Racah problem for the positive-discrete series of irreducible representations of $\mathfrak{su}(1,1)$ is reviewed. The Bargmann realization of these representations is given and the definition of the $6j$-symbols of the algebra in terms of overlap coefficients between eigenbases associated to intermediate Casimir operators is provided.
\subsection{Positive-discrete series representations and Bargmann realization of $\mathfrak{su}(1,1)$}
The $\mathfrak{su}(1,1)$ algebra has three generators $K_{\pm}$, $K_0$ as its basis elements. These generators obey the commutation relations
\begin{align}
\label{SU}
[K_0,K_{\pm}]=\pm K_{\pm},\qquad [K_{-},K_{+}]=2K_0.
\end{align}
The Casimir operator $\mathcal{Q}$, which commutes with all $\mathfrak{su}(1,1)$ elements, is given by
\begin{align}
\label{Casimir-SU}
\mathcal{Q}=K_0^2-K_0-K_{+}K_{-}.
\end{align}
We shall here be concerned with irreducible representations of \eqref{SU} belonging to the positive-discrete series. These representations are labeled by a positive number $\nu$ and can be defined by the following actions of the $\mathfrak{su}(1,1)$ generators on basis vectors $e_{n}$, $n\in \mathbb{N}$ :
\begin{align}
\label{Actions}
K_0e_{n}=(n+\nu)e_{n},\qquad K_{-}e_{n}=ne_{n-1},\qquad K_{+}e_{n}=(n+2\nu)e_{n+1}.
\end{align}
One can realize the positive-discrete representations on the space of holomorphic functions of a single variable $x$. In this realization, the $\mathfrak{su}(1,1)$ generators take the form \cite{Perelomov-1986}
\begin{align}
\label{Bargmann}
K_0=x\pd_{x}+\nu,\quad K_{-}=\pd_{x},\quad K_{+}=x^2\pd_{x}+2\nu x,
\end{align}
and the Casimir operator is a multiple of the identity
\begin{align*}
\mathcal{Q}=\nu(\nu-1).
\end{align*}
It is easily seen that on the monomial basis 
\begin{align}
\label{Monomials}
e_{n}(x)=x^{n},
\end{align}
with $n\in \mathbb{N}$, the actions \eqref{Actions} are recovered.
\subsection{Addition schemes for three $\mathfrak{su}(1,1)$ algebras}
Consider three mutually commuting sets\footnote{Here the symbol $\{\}$ for sets should not be confused with the anticommutator.} of $\mathfrak{su}(1,1)$ generators $\mathbf{K}^{(i)}=\big\{K_0^{(i)},\, K_{\pm}^{(i)}\big\}$, $i=1,2,3$, with $[\mathbf{K}^{(i)},\mathbf{K}^{(j)}]=0$ for $i\neq j$. These sets of generators can be combined by addition to form four additional ones  $\mathbf{K}^{(12)}$, $\mathbf{K}^{(23)}$, $\mathbf{K}^{(31)}$ and $\mathbf{K}^{(4)}$ defined by
\begin{align}
\label{Intermediate-Algebra}
\mathbf{K}^{(ij)}=\big\{K_0^{(ij)}\equiv K_0^{(i)}+K_0^{(j)},\,K_{\pm}^{(ij)}\equiv K_{\pm}^{(i)}+K_{\pm}^{(j)} \big\},
\end{align}
and
\begin{align}
\label{Full-Algebra}
\mathbf{K}^{(4)}=\big\{K_0^{(4)}\equiv K_0^{(1)}+K_0^{(2)}+K_0^{(3)},\,K_{\pm}^{(4)}\equiv K_{\pm}^{(1)}+K_{\pm}^{(2)}+K_{\pm}^{(3)} \big\}.
\end{align}
The Casimir operators associated to \eqref{Intermediate-Algebra} and \eqref{Full-Algebra} have the expressions:
\begin{subequations}
\begin{align}
\label{Intermediate-Casimir}
\mathcal{Q}^{(ij)}&=[K_0^{(ij)}]^2-K_0^{(ij)}-K_{+}^{(ij)}K_{-}^{(ij)},
\\
\label{Full-Casimir}
\mathcal{Q}^{(4)}&=[K_0^{(4)}]^2-K_0^{(4)}-K_{+}^{(4)}K_{-}^{(4)}.
\end{align}
\end{subequations}
The Casimir operators \eqref{Intermediate-Casimir} will be referred to as the ``intermediate Casimirs'' whereas the operator \eqref{Full-Casimir} will be referred to as the ``full Casimir''. The intermediate Casimir operators $\mathcal{Q}^{(ij)}$ and the full Casimir operator $\mathcal{Q}^{(4)}$ are not independent. Indeed, an elementary calculation shows that
\begin{align}
\label{elem}
\mathcal{Q}^{(4)}=\mathcal{Q}^{(12)}+\mathcal{Q}^{(23)}+\mathcal{Q}^{(31)}-\mathcal{Q}^{(1)}-\mathcal{Q}^{(2)}-\mathcal{Q}^{(3)},
\end{align}
where $\mathcal{Q}^{(j)}$, $j=1,2,3$, are the Casimir operators \eqref{Casimir-SU} associated to each set $\mathbf{K}^{(i)}$. As is easily verified, the Casimir operators \eqref{Intermediate-Casimir} commute with the Casimir operators $\mathcal{Q}^{(i)}$ and with the total Casimir operator $\mathcal{Q}^{(4)}$, but do not commute amongst themselves. The full Casimir operator $\mathcal{Q}^{(4)}$ commutes with both the intermediate Casimirs $\mathcal{Q}^{(ij)}$ and with the individual Casimirs $\mathcal{Q}^{(i)}$.
\subsection{$6j$-symbols}
The $6j$-symbols, also known as the Racah coefficients, arise in the following situation. Consider three irreducible representations of the positive-discrete series labeled by the parameters $\nu_i$, $i=1,2,3$ associated to the eigenvalues $\nu_i(\nu_i-1)$ of the individual Casimir operators $\mathcal{Q}^{(i)}$. In this case, the representation parameters $\nu_{ij}$ associated to the eigenvalues $\nu_{ij}(\nu_{ij}-1)$ of the intermediate Casimir operators $\mathcal{Q}^{(ij)}$ have the form $\nu_{ij}=\nu_i+\nu_j+n_{ij}$, where the $n_{ij}$ are non-negative integers. Furthermore, the possible values for the representation parameter $\nu_4$ associated to the eigenvalues $\nu_4(\nu_4-1)$ of the full Casimir operator $\mathcal{Q}^{(4)}$ are given by $\nu_4=\nu_{12}+\nu_3+\ell=\nu_1+\nu_{23}+m=\nu_1+\nu_2+\nu_3+k$, where $m$, $\ell$ and $k$ are non-negative integers. For details, the reader can consult \cite{VDJ-2003,Rosengren-2007-06}.

For a given value of the total Casimir parameter $\nu_4$, one has a finite-dimensional space on which the pair of (non-commuting) operators $\mathcal{Q}^{(12)}$, $\mathcal{Q}^{(23)}$ act. Each of these operators has a set of eigenvectors $\{\phi_{n_{12}}\}_{n_{12}=0}^{K}$, $\{\chi_{n_{23}}\}_{n_{23}=0}^{K}$  such that
\begin{align}
\label{1}
\mathcal{Q}^{(12)}\phi_{n_{12}}=\nu_{12}(\nu_{12}-1)\phi_{n_{12}},\qquad \mathcal{Q}^{(23)}\chi_{n_{23}}=\nu_{23}(\nu_{23}-1)\chi_{n_{23}},
\end{align}
where $\nu_{12}=\nu_{1}+\nu_{2}+n_{12}$ and $\nu_{23}=\nu_{1}+\nu_{2}+n_{23}$. Both sets of basis vectors $\{\phi_{n_{12}}\}$, $\{\chi_{n_{23}}\}$ are eigenvectors of the Casimir operators $\mathcal{Q}^{(i)}$, with $i=1,\ldots,4$. The $6j$-symbols are the overlap coefficients $W_{n_{12}, n_{23}}$ between the two bases
\begin{align}
\label{2}
\phi_{n_{12}}=\sum_{n_{23}=0}^{K}W_{n_{12},n_{23}}\,\chi_{n_{23}}.
\end{align}
The dimension $K+1$ of the space can be evaluated straightforwardly in terms of the representation parameters $\nu_i$, $i=1,\ldots,4$. If $\nu_4=\nu_1+\nu_2+\nu_3+M$, then it follows from the above considerations that $\nu_{12}$ can take the $M+1$ possible values $\nu_{12}\in\{\nu_1+\nu_2,\,\nu_1+\nu_2+1,\ldots,\,\nu_1+\nu_2+M\}$ while $\nu_{23}$ can take the $M+1$ values $\nu_{23}\in\{\nu_2+\nu_3,\,\nu_2+\nu_3+1,\ldots,\nu_2+\nu_3+M\}$. Thus, for a fixed value $\nu_4$, the dimension $K+1$ of the space is determined by the value $K=\nu_4-\nu_1-\nu_2-\nu_3$ with $K\in\mathbb{N}$. Note that one can also consider (non-standard) $6j$-symbols for the pairs of operators $\mathcal{Q}^{(23)}$, $\mathcal{Q}^{(31)}$ and $\mathcal{Q}^{(12)}$, $\mathcal{Q}^{(31)}$.
\section{The Racah algebra and the Racah problem}
In this section, it is shown that the Racah algebra is behind the Racah problem for $\mathfrak{su}(1,1)$. This result follows from the determination of the commutation relations satisfied by the intermediate Casimir operators of the Racah problem. The generators satisfying the $\mathbb{Z}_3$-symmetric Racah relations are also exhibited.
\subsection{Racah Algebra}
Let $A$, $B$ be expressed as follows in terms the intermediate Casimir operators:
\begin{align}
\label{Rea}
A=-\mathcal{Q}^{(12)}/2,\qquad B=-\mathcal{Q}^{(23)}/2,
\end{align}
and define $C=[A,B]$. In the Racah problem, the Casimir operators $\mathcal{Q}^{(i)}$, $i=1,\ldots, 4$ act as multiples of the identity and hence they can be replaced by constants $Q^{(i)}=\lambda_i$, where $\lambda_i=\nu_i(\nu_i-1)$. A direct computation shows that the operators $A$, $B$, together with their commutator $C$, satisfy the defining relations of the Racah algebra
\begin{subequations}
\label{Commu}
\begin{align}
[A,B]&=C,
\\
[B,C]&=B^2+\{A,B\}+\delta B+\epsilon_1,
\\
[C,A]&=A^2+\{A,B\}+\delta A+\epsilon_2,
\end{align}
\end{subequations}
where
\begin{align}
\delta=(\lambda_1+\lambda_2+\lambda_3+\lambda_4)/2,\qquad \epsilon_1=(\lambda_1-\lambda_4)(\lambda_2-\lambda_3)/4,\qquad \epsilon_2=(\lambda_1-\lambda_2)(\lambda_4-\lambda_3)/4.
\end{align}
The commutation relations \eqref{Commu} are most easily verified using the Bargmann realization \eqref{Bargmann} in three variables $x$, $y$ and $z$ but hold regardless of the representation. It is seen that the relations \eqref{Commu} are exactly the defining relations \eqref{Def-Rel} of the Racah algebra. In the realization \eqref{Rea}, it is directly checked that the Casimir operator \eqref{Casimir-Racah} takes the value
\begin{align}
\label{Casimir-Value}
Q=\frac{1}{4}\big[(\lambda_1-\lambda_2+\lambda_3-\lambda_4)(\lambda_1\lambda_3-\lambda_2\lambda_4)-\lambda_1\lambda_2-\lambda_2\lambda_3-\lambda_3\lambda_4-\lambda_4\lambda_1\big].
\end{align}
It is easy to see that any pair of intermediate Casimir operators $(\mathcal{Q}^{(ij)},\mathcal{Q}^{(k\ell)})$ will satisfy the relations \eqref{Commu}. Therefore the intermediate Casimir operators $(\mathcal{Q}^{(12)},\mathcal{Q}^{(23)},\mathcal{Q}^{(31)})$ realize a \emph{Leonard Triple} \cite{Curtin-2007-07}.
\subsection{$\mathbb{Z}_3$-symmetric presentation of the Racah problem}
Let $X$, $Y$, $Z$ be defined as follows in terms of the intermediate Casimir operators of the $\mathfrak{su}(1,1)$ Racah problem:
\begin{align}
\label{XYZ}
X=\mathcal{Q}^{(12)}-\mathcal{Q}^{(1)},\qquad Y=\mathcal{Q}^{(23)}-\mathcal{Q}^{(2)},\qquad Z=\mathcal{Q}^{(31)}-\mathcal{Q}^{(3)}.
\end{align}
In view of \eqref{elem}, one has
\begin{align*}
X+Y+Z=\mathcal{Q}^{(4)}=\lambda_4.
\end{align*}
Upon comparing \eqref{XYZ} with \eqref{XYZ-2}, it follows that the operators \eqref{XYZ} obey the $\mathbb{Z}_3$-symmetric Racah relations. It is also seen that the value of the Casimir operator for the Racah algebra \eqref{Casimir-Value} also possesses this symmetry. It is easy to understand the origin of the $\mathbb{Z}_3$ symmetry in this context: it corresponds to the $\mathbb{Z}_3$ freedom in permuting the three $\mathfrak{su}(1,1)$ representations $\mathbf{K}^{(i)}$. Hence the Racah problem for $\mathfrak{su}(1,1)$ is intrinsically $\mathbb{Z}_3$-symmetric. We shall now consider the Racah problem in the Bargmann picture.

\section{Sturm--Liouville model for the Racah algebra}
In this section, the Racah problem for $\mathfrak{su}(1,1)$ is considered in the Bargmann representation. It is shown that in this picture, the determination of the Racah coefficients is equivalent to obtaining the overlap coefficients between solutions to pairs of hypergeometric Sturm--Liouville problems. This gives a realization of the Racah algebra in terms of differential operators of a single variable.
\subsection{Racah problem in the Bargmann picture}
Consider the problem of determining the Racah coefficients as defined by the set of equations \eqref{1}, \eqref{2}. It is clear from this definition that one can arbitrarily choose the value of the projection operator $K_0^{(4)}$ on the involved basis vectors. Let $\psi$ be an eigenvector of every Casimir operator $\mathcal{Q}^{(i)}$, $i=1,\ldots,4$, with the minimal value of this projection. In view of \eqref{Actions}, this is means that
\begin{align}
\label{Conditions-1}
K_0^{(4)}\psi=(\nu_1+\nu_2+\nu_3+M)\psi,\qquad K_{-}^{(4)}\psi=(K_-^{(1)}+K_-^{(2)}+K_-^{(3)})\psi=0,
\end{align}
where $M$ is a non-negative integer. In the Bargmann realization \eqref{Bargmann}, $\psi=\psi(x,y,z)$ and it is seen from \eqref{Monomials} and \eqref{Conditions-1} that $\psi(x,y,z)$ can be expressed as a polynomial in the variables $x$, $y$, $z$ of total degree $M$. The conditions \eqref{Conditions-1} translate into
\begin{align}
\label{Conditions-2}
(x\pd_{x}+y\pd_{y}+z\pd_{z})\psi(x,y,z)=M \psi(x,y,z),\qquad (\pd_x+\pd_y+\pd_z)\psi(x,y,z)=0.
\end{align}
By Euler's homogeneous function theorem, the first condition of \eqref{Conditions-2} implies that $\psi(x,y,z)$ is homogeneous of degree $M$, which means that
\begin{align*}
\psi(\alpha x,\alpha y,\alpha z)=\alpha^{M}\psi(x,y,z).
\end{align*}
The second condition of \eqref{Conditions-2} implies that $\psi(x,y,z)$ can only depend on the relative variables $(x-y)$ and $(z-y)$. Hence it follows that the most general expression for $\psi(x,y,z)$ is 
\begin{align*}
\psi(x,y,z)=(z-y)^{M} \Phi\left(\frac{x-y}{z-y}\right),
\end{align*}
where $\Phi(u)$ is a polynomial in $u$ of maximal degree $M$. It is seen that the action of the intermediate Casimir operators is given by
\begin{subequations}
\begin{gather}
\mathcal{Q}^{(12)}\psi(x,y,z)=(z-y)^{M}\mathcal{S}_{12}\Phi(u),\qquad
\mathcal{Q}^{(23)}\psi(x,y,z)=(z-y)^{M}\mathcal{S}_{23}\Phi(u),
\\
\mathcal{Q}^{(31)}\psi(x,y,z)=(z-y)^{M}\mathcal{S}_{31}\Phi(u),
\end{gather}
\end{subequations}
where the one-variable operators $\mathcal{S}_{ij}$ are given by
\begin{align}
\begin{aligned}
\label{Hyper}
\mathcal{S}_{12}&=u^2(1-u)\pd_{u}^2+u\big[(M-1-2\nu_1)u+2(\nu_1+\nu_2)\big]\pd_{u}+2M\nu_1 u+(\nu_1+\nu_2)(\nu_1+\nu_2-1),
\\
\mathcal{S}_{23}&=u(u-1)\pd_u^2+\big[2(1-M-\nu_2-\nu_3)u+(M-1+2\nu_3)\big]\pd_u+(M+\nu_2+\nu_3)(M+\nu_2+\nu_3-1),
\\
\mathcal{S}_{31}&=u(u-1)^2\pd_u^2+(1-u)\big[(M-1-2\nu_1)u+1-M-2\nu_3\big]\pd_u +2M\nu_1(1-u)+(\nu_1+\nu_3)(\nu_1+\nu_3-1).
\end{aligned}
\end{align}
\normalsize
It is elementary to verify that the operators $\mathcal{S}_{ij}$ preserve the space of polynomials of maximal degree $M$. The operator $\mathcal{S}_{23}$ is a standard hypergeometric operator while $\mathcal{S}_{12}$ and $\mathcal{S}_{13}$ can be reduced to hypergeometric operators by appropriate changes of variables.

Returning to the Racah problem for the intermediate Casimir operators $\mathcal{Q}_{12}$ and $\mathcal{Q}_{23}$, it follows from the above that the equations \eqref{1} are equivalent to the pair of Sturm--Liouville problems
\begin{align}
\label{3}
\mathcal{S}_{12}\Phi^{(12)}(u)=\nu_{12}(\nu_{12}-1)\Phi^{(12)}(u),\qquad \mathcal{S}_{23}\Phi^{(23)}(u)=\nu_{23}(\nu_{23}-1)\Phi^{(23)}(u),
\end{align}
where $\Phi^{(12)}(u)$ and $\Phi^{(23)}(u)$ are required to be polynomials in $u$ of degree not higher than $M$. In this picture, the Racah decomposition \eqref{2} becomes
\begin{align*}
\Phi^{(12)}(u)=\sum_{n_{23}=0}^{M}W_{n_{12},n_{23}}\Phi^{(23)}(u),
\end{align*}
where it is assumed that $\nu_{12}=n_{12}+\nu_1+\nu_2$ and $\nu_{23}=n_{23}+\nu_2+\nu_3$. The explicit solutions to the Sturm--Liouville equations \eqref{3} can be found in terms of Gauss hypergeometric functions. Indeed, consider the eigenvalue equation
\begin{align*}
\mathcal{S}_{12}\Phi^{(12)}(u)=\nu_{12}(\nu_{12}-1)\Phi^{(12)}(u),
\end{align*}
with $\nu_{12}=n_{12}+\nu_1+\nu_2$. Then using \eqref{Hyper}, it is directly verified that the polynomial solutions for $\Phi^{(12)}(u)$, up to an inessential constant factor, are given by
\begin{align*}
\Phi^{(12)}(u)=u^{n_{12}}\,\pFq{2}{1}{n_{12}-M,n_{12}+2\nu_1}{2n_{12}+2\nu_1+2\nu_2}{u},
\end{align*}
where ${}_2F_1$ is the Gauss hypergeometric function
\begin{align}
\label{2F1}
\pFq{2}{1}{a,b}{c}{z}=\sum_{i=0}^{\infty}\frac{(a)_i(b)_i}{(c)_i}\frac{z^{i}}{i!},
\end{align}
and where $(a)_i=(a)(a+1)\cdots (a+i-1)$ stands for the Pochhammer symbol. Using \eqref{2F1}, the solution for $\Phi^{(12)}(u)$ can also be presented in the form
\begin{align*}
\Phi^{(12)}(u)=u^{M}\;\pFq{2}{1}{n_{12}-M,1-M-n_{12}-2\nu_1-2\nu_2}{1-M-2\nu_1}{\frac{1}{u}}.
\end{align*}
Proceeding similarly for $\Phi^{(23)}(u)$ and $\Phi^{(31)}(u)$, one finds that 
\begin{align*}
\Phi^{(23)}(u)&=\pFq{2}{1}{n_{23}-M,1-M-n_{23}-2\nu_2-2\nu_3}{1-M-2\nu_3}{u},
\\
\Phi^{(31)}(u)&=(1-u)^{M}\;\pFq{2}{1}{n_{31}-M,1-M-n_{31}-2\nu_3-2\nu_1}{1-M-2\nu_1}{\frac{1}{1-u}},
\end{align*}
where $n_{31}=\nu_{31}-\nu_1-\nu_3$. Thus, in the Bargmann picture, the Racah coefficients $W_{n_{12},n_{23}}$ occur as overlap coefficients between the solutions of a pair of Sturm--Liouville problems.
\subsection{One-variable realization of the Racah algebra and equitable presentation}
The reduction from a three-variable model to a one-variable model for the Racah problem in the Bargmann picture can be used to exhibit a one-variable realization of the Racah algebra and its equitable presentation. Indeed, it is directly checked that the one-variable operators
\begin{align*}
\kappa_1=-\mathcal{S}_{12}/2,\qquad \kappa_2=-\mathcal{S}_{23}/2,
\end{align*}
together with their commutator $\kappa_3=[\kappa_1,\kappa_2]$, realize the Racah algebra \eqref{Commu} under the identification $\kappa_1=A$, $\kappa_2=B$. Furthermore, since $M=\nu_4-\nu_1-\nu_2-\nu_3$, one sees that the relation
\begin{align*}
\mathcal{S}_{12}+\mathcal{S}_{23}+\mathcal{S}_{31}=\nu_4(\nu_4-1)+\nu_3(\nu_3-1)+\nu_2(\nu_2-1)+\nu_1(\nu_1-1),
\end{align*}
holds and one finds that the operators
\begin{align}
\label{XYZ-3}
X=\mathcal{S}_{12}-\nu_1(\nu_1-1),\qquad Y=\mathcal{S}_{23}-\nu_2(\nu_2-1),\qquad Z=\mathcal{S}_{31}-\nu_3(\nu_3-1),
\end{align}
are related by $X+Y+Z=\lambda_4$ and satisfy the $\mathbb{Z}_3$-symmetric Racah relations \eqref{Z3-Symmetric-Racah} with $\lambda_i=\nu_i(\nu_i-1)$.
\section{The Racah algebra and the equitable $\mathfrak{su}(2)$ algebra}
In the previous section, the reduction from a three-variable to a one-variable model for the Racah problem was performed and led to a one-variable realization of the Racah algebra. In this section, another interpretation of the operators $\mathcal{S}_{ij}$ in terms of elements in the enveloping algebra of $\mathfrak{su}(2)$ algebra is presented. Using this interpretation, the finite-dimensional irreducible representations of $\mathfrak{su}(2)$ are used to define irreducible representations of the Racah algebra.

\subsection{Equitable presentation of the $\mathfrak{su}(2)$ algebra}
The $\mathfrak{su}(2)$ algebra consists of three generators $J_0$, $J_{\pm}$ satisfying the commutation relations
\begin{align*}
[J_0,J_{\pm}]=\pm J_{\pm},\qquad [J_{+},J_{-}]=2J_0.
\end{align*}
The Casimir operator for $\mathfrak{su}(2)$, denoted $\Delta$, is given by
\begin{align}
\label{SU-2-Casimir}
\Delta=J_0^{2}-J_0-J_{+}J_{-}.
\end{align}
All unitary irreducible representations of $\mathfrak{su}(2)$ are finite-dimensional. In these representations of dimension $2j+1$, the Casimir operator takes the value $j(j+1)$ with $j\in\{0,\,1/2,\,1,\,3/2,\ldots\}$. The $\mathfrak{su}(2)$ algebra has the Bargmann realization
\begin{align}
\label{4}
J_{-}=-\pd_u,\qquad J_{+}=u^2\pd_u-2ju,\qquad J_0=u\pd_u-j.
\end{align}
In the realization \eqref{4}, one has $\Delta=j(j+1)$ for the Casimir operator. There exists another presentation of the $\mathfrak{su}(2)$ algebra known as the \emph{equitable} presentation \cite{Hartwig-2006-10}. The equitable basis is defined by
\begin{align}
\label{Equitable}
E_1=2(J_{+}-J_0),\qquad E_2=-2(J_-+J_0),\qquad E_3=2J_0.
\end{align}
in terms of which the commutation relations read
\begin{align*}
[E_i,E_{j}]=2(E_i+E_{j}),
\end{align*}
where $(ij)\in\{(12),(23),(31)\}$.
\subsection{Equitable Racah operators from equitable $\mathfrak{su}(2)$ generators}
Let us explain how the equitable Racah relations can be realized with quadratic elements in the $\mathfrak{su}(2)$ algebra; this observation has been made in \cite{Gao-2013-06}. We have already seen in \eqref{XYZ-3} that the operators $X$, $Y$, $Z$ of the $\mathbb{Z}_3$-symmetric presentation of the Racah algebra \eqref{Z3-Symmetric-Racah} can be realized by one-variable differential operators. Let $\mathcal{G}_i$, $i=1,2,3$, be the following quadratic elements in the equitable $\mathfrak{su}(2)$ generators:
\begin{subequations}
\label{SU-Rea}
\begin{align}
\mathcal{G}_1=-\frac{1}{8}\{E_1,E_3\}+\frac{\nu_2-\nu_1}{2}(E_3+E_1)+\frac{1-M-2\nu_1-2\nu_2}{4}(E_1-E_3)+\frac{(M+2\nu_2)(M+4\nu_1+2\nu_2-2)}{4},
\\
\mathcal{G}_2=-\frac{1}{8}\{E_2,E_3\}+\frac{\nu_3-\nu_2}{2}(E_2+E_3)+\frac{1-M-2\nu_2-2\nu_3}{4}(E_3-E_2)+\frac{(M+2\nu_3)(M+4\nu_2+2\nu_3-2)}{4},
\\
\mathcal{G}_3=-\frac{1}{8}\{E_1,E_2\}+\frac{\nu_1-\nu_3}{2}(E_1+E_2)+\frac{1-M-2\nu_1-2\nu_3}{4}(E_2-E_1)+\frac{(M+2\nu_1)(M+4\nu_3+2\nu_1-2)}{4}.
\end{align}
\end{subequations}
\normalsize
Then one has $\mathcal{G}_1+\mathcal{G}_2+\mathcal{G}_3=\lambda_4=\nu_4(\nu_4-1)$ when $M=\nu_4-\nu_1-\nu_2-\nu_3$. When the Bargmann realization \eqref{4} is used with $j=M/2$, the operators $\mathcal{G}_i$ are identified with the one-variable realizations $\mathcal{S}_{ij}$ of the intermediate Casimir operators through
\begin{align}
\label{Iden}
\mathcal{G}_{1}=\mathcal{S}_{12}-\nu_1(\nu_1-1)=X,\qquad \mathcal{G}_{2}=\mathcal{S}_{23}-\nu_2(\nu_2-1)=Y,\qquad \mathcal{G}_{3}=\mathcal{S}_{31}-\nu_3(\nu_3-1)=Z.
\end{align}
Hence the quadratic elements $\mathcal{G}_i$ in the $\mathfrak{su}(2)$ generators realize the $\mathbb{Z}_3$-symmetric Racah relations \eqref{Z3-Symmetric-Racah}.
\subsection{Racah algebra representations from $\mathfrak{su}(2)$ modules}
The standard basis for the irreducible representations of $\mathfrak{su}(2)$ and the realization \eqref{SU-Rea} of the Racah algebra can be used to construct finite-dimensional representations of the Racah algebra. Let $e_{n}$, $n=0,1,\ldots, M$, denote the canonical basis vectors for the $M+1$-dimensional irreducible representations of $\mathfrak{su}(2)$. These representations are defined by the actions
\begin{align}
\label{Action-1}
J_0e_{n}=(n-M/2)e_{n},\qquad J_{+}e_{n}=\sqrt{(n+1)(M-n)}e_{n+1},\qquad J_{-}e_{n}=\sqrt{n(M-n+1)}e_{n-1}.
\end{align}
In this basis, the equitable generators \eqref{Equitable} act in the following way:
\begin{subequations}
\label{Dompe}
\begin{align}
E_1e_{n}&=(M-2n)e_{n}+2\sqrt{(n+1)(M-n)}e_{n+1},
\\
E_2e_{n}&=-2\sqrt{n(M-n+1)}e_{n-1}+(M-2n)e_{n},
\\
E_3e_{n}&=(2n-M)e_{n}.
\end{align}
\end{subequations}
Let $A$ and $B$ be defined as
\begin{align}
\label{Rea-2}
A=-\frac{1}{2}\mathcal{G}_1-\nu_1(\nu_1-1)/2,\qquad B=-\frac{1}{2}\mathcal{G}_2-\nu_2(\nu_2-1)/2,
\end{align}
where $\mathcal{G}_1$, $\mathcal{G}_2$ are as in \eqref{SU-Rea}. It follows from \eqref{Iden} that the operators $A$ and $B$ realize the Racah algebra \eqref{Commu} with $\lambda_i=\nu_i(\nu_i-1)$ and $\nu_4=M+\nu_1+\nu_2+\nu_3$. A direct computation using \eqref{Dompe} shows that in the basis $e_{n}$, the operators $A$ and $B$ have the actions

\begin{subequations}
\label{Action-3}
\begin{align}
A\,e_{n}&=\lambda_n^{(A)}e_{n}+\frac{1}{2}(n+2\nu_1)\sqrt{(n+1)(M-n)}\,e_{n+1},
\\
B\,e_{n}&=\lambda_{n}^{(B)}e_{n}+\frac{1}{2}(M-n+2\nu_3)\sqrt{n(M-n+1)}\,e_{n-1},
\end{align}
\end{subequations}
where
\begin{subequations}
\label{Eigen}
\begin{align}
\lambda_{n}^{(A)}&=-(n+\nu_1+\nu_2)(n+\nu_1+\nu_2-1)/2,
\\
\lambda_{n}^{(B)}&=-(M-n+\nu_2+\nu_3)(M-n+\nu_2+\nu_3-1)/2.
\end{align}
\end{subequations}
Since $A$ and $B$ act in a bidiagonal fashion, the expression \eqref{Eigen} are the eigenvalues of $A$ and $B$ in this representation. In the generic case, the $(M+1)$-dimensional representations of the Racah algebra defined by \eqref{Action-3} are clearly irreducible. It is convenient at this point to introduce another basis spanned by the basis vectors $\widetilde{e}_n$ which are defined by
\begin{align}
e_{n}=\sqrt{\frac{(-1)^{n}}{n!(-M)_{n}}}\frac{2^{n}}{(2\nu_1)_{n}}\widetilde{e}_{n}.
\end{align}
On the basis vectors $\widetilde{e}_{n}$, the actions \eqref{Action-3} are
\begin{subequations}
\label{Action-4}
\begin{align}
A\widetilde{e}_{n}&=\lambda_n^{(A)}\widetilde{e}_{n}+\widetilde{e}_{n+1},
\\
B\widetilde{e}_{n}&=\lambda_{n}^{(B)}\widetilde{e}_{n}+\varphi_{n}\widetilde{e}_{n-1},
\end{align}
\end{subequations}
where
\begin{align}
\varphi_{n}=n(M-n+1)(n+2\nu_1-1)(M-n+2\nu_3)/4.
\end{align}
From \eqref{Action-4}, it is seen that the basis spanned by the vectors $\widetilde{e}_n$ corresponds to the UD-LD basis for Leonard pairs studied by Terwilliger in \cite{Terwilliger-2005-09}. See also \cite{Alnajjar-2010-08} for realizations of Leonard pairs using the equitable generators of $\mathfrak{sl}_2$.

\section{Conclusion} 
In this paper, we have established the correspondence between two frameworks for the realization of the Racah algebra: the one in which the Racah algebra is realized by the intermediate Casimir operators arising in the combination of three $\mathfrak{su}(1,1)$ representations of the positive-discrete series and the one where the Racah is realized in terms of quadratic elements in the enveloping algbera of $\mathfrak{su}(2)$. We have also exhibited how the $\mathbb{Z}_3$-symmetric, or equitable, presentation of the Racah algebra arises in the context of the Racah problem for $\mathfrak{su}(1,1)$.

In \cite{Genest-2012-08,Genest-2012}, it was shown that the Bannai-Ito (BI) algebra is the algebraic structure behind the Racah problem for the $sl_{-1}(2)$ algebra and a $\mathbb{Z}_3$-symmetric presentation of the BI algebra was offered. In view of the results presented here, it would be of interest to perform the reduction of the number of variables in the $sl_{-1}(2)$ Racah problem to obtain a one-variable realization of the Bannai-Ito algebra and to identify in this case what is the algebraic structure that plays a role analogous to the one played here by $\mathfrak{su}(2)$.
\section*{Aknowledgments}
The authors thank S. Gao for making reference \cite{Gao-2013-06} available to them in proofs and are grateful to P. Terwilliger for stimulating discussions. LV benefited from the hospitality of the Donetsk Institute for Physics and Technology while this work was underway. VXG holds a Alexander-Graham-Bell fellowship from the Natural Science and Engineering Research Council of Canada (NSERC). The research of LV is supported in part by NSERC.
\footnotesize
\begin{multicols}{2}

\end{multicols}
\end{document}